\documentclass{vldb}
\usepackage{graphicx}
\usepackage{balance} 
\usepackage{url}
\usepackage{graphicx}
\usepackage{xcolor}
\usepackage{bera}
\usepackage{listings}

\colorlet{punct}{red!60!black}
\definecolor{background}{HTML}{EEEEEE}
\definecolor{delim}{RGB}{20,105,176}
\colorlet{numb}{magenta!60!black}
\lstdefinelanguage{json}{
    basicstyle=\normalfont\ttfamily,
    numbers=left,
    numberstyle=\scriptsize,
    stepnumber=1,
    numbersep=8pt,
    showstringspaces=false,
    breaklines=true,
    frame=lines,
    backgroundcolor=\color{background},
    literate=
     *{0}{{{\color{numb}0}}}{1}
      {1}{{{\color{numb}1}}}{1}
      {2}{{{\color{numb}2}}}{1}
      {3}{{{\color{numb}3}}}{1}
      {4}{{{\color{numb}4}}}{1}
      {5}{{{\color{numb}5}}}{1}
      {6}{{{\color{numb}6}}}{1}
      {7}{{{\color{numb}7}}}{1}
      {8}{{{\color{numb}8}}}{1}
      {9}{{{\color{numb}9}}}{1}
      {:}{{{\color{punct}{:}}}}{1}
      {,}{{{\color{punct}{,}}}}{1}
      {\{}{{{\color{delim}{\{}}}}{1}
      {\}}{{{\color{delim}{\}}}}}{1}
      {[}{{{\color{delim}{[}}}}{1}
      {]}{{{\color{delim}{]}}}}{1},
}

\begin{document}

\title{Providing better confidentiality and authentication on the Internet using Namecoin and MinimaLT}

\numberofauthors{1}
\author{
\alignauthor
Frederic Jacobs\\
\affaddr{www.fredericjacobs.com}\\
\email{me@fredericjacobs.com}
}

\maketitle

\begin{abstract}
In this paper, we introduce a duo of improvements for the Internet that would lead to better security. The authentication model on the Internet is broken and TLS connections have a considerable overhead. We try to address those issues with changes in both the application layer, discussing a replacement for the DNS system, and in the transport layer, a drop-in replacement for TCP built on top of UDP so that it can run on today's internet infrastructure.

\end{abstract}

\section{Introduction}

\subsection{Defining user privacy}

The solutions brought forward in this paper are attempts to fix \emph{confidentiality and authentication} on the Internet. \emph{Anonymity is not provided.} Although, an attacker could still get a significant amount of metadata because IP headers are left unchanged, MinimaLT does provide a similar traffic analysis protection as IPSec by using tunnels.

Unfortunately, because MinimaLT runs over UDP it is not possible to use it through the Tor network.\footnote{If MinimaLT proves to be a safer and faster alternative to TLS, I imagine that the Tor project would look into implementing it to speed up the network and make relay connections safer.}

\subsection{Motivation}

When the Internet was designed at DARPA, the primary goal was to design a system that could provide interconnection between multiple computers. The Web then came by with the motivation to be able to freely exchange information with anyone. But, over time, people started trusting the internet more and more and started sharing more personal information over it. The threat model implied that you had to trust the entities running the infrastructure of the internet, but the latter was never designed to be run by so many of them. Over time, attempts such as TLS and DNSSec were introduced to secure inherently insecure protocols. In this paper, we will not attempt to fix the current protocols that have a huge overhead but we will try instead to propose better alternatives.

\section{Domain names and authenticity}

Today, if we want to load a page from \emph{www.facebook.com}, our computer will have to first get the Domain Name System record matching that domain. DNS was designed in a hierarchical way and TLD registrations are handled by a single organisation, the ICANN.

\emph{So what is wrong with DNS?}

When the original Domain Name System was designed, it did not include security; instead it was designed to be a scalable distributed system. DNS requests could thus be spoofed and fake DNS query responses could be served to the clients. Therefore, DNSSEC, a security extension of DNS, attempted to prevent those types of attacks by introducing DNS zone signing. DNS zone signing uses a chain of trust to sign entries hierarchically,  meaning that the ICANN has a key that is known as the root anchor which is the starting point of a trust chain.

The fact that even today, DNSSEC is having issues being deployed at a larger scale has a lot to do with its complicated design. Getting DNSSEC right is hard and it leads to the centralisation of the internet Domain Name System because few registrars are deploying DNSSEC and other registrars are still requesting their signing keys. Hierarchical trust structures require us to blindly believe that the root keys are not compromised and bring us back to a feudal system where we need to beg some lords (here companies) for protection to get the keys.

The ICANN has been taking worrying measures that remove the users' ability to register domains anonymously\cite{icannabuse}. Our systems shouldn't be designed in such a way that a change in policy would make their security obsolete. 

Hence, we want to design a distributed system where anyone can register a domain with a fast and efficient registration process.

But how can we verify authenticity? 
Even if our Domain Name System returns the right IP address, how do we know for sure that we are establishing a connection with the client we want ? Today, we are using yet another hierarchical system to verify authenticity, namely \emph{SSL certificates}. This means that in addition to trusting ICANN, we will have to trust hundreds of Root Certificate Authorities that are shipped with our browsers.\cite{mozillaSSL}

If only one of those 100s of CAs gets compromised, it could result in the man-in-the-middling of any website without any warning since a root certification authority can generate a fake valid certificate for said website.

Certificate Authorities getting compromised has already happened. For instance, the Dutch Certificate Authority DigiNotar's hack \cite{diginotarHack} enabled fake certificates to be delivered in order to target Iranian citizens.

Now that we are convinced that the hierarchical trust model of the internet is broken, one might wonder what measures have already been taken to fix authentication on the Internet?

\section{Existing attempts to address authentication}

\subsubsection{Certificate Pinning}

The Chrome Security team led the way by implementing certificate pinning. Certificate pinning is an effective measure to counter man-in-the-middle attacks in today's internet. Certificate pinning works by shipping \emph{pins}\footnote{A \emph{pin} is a (domain name, certificate fingerprint) pair} in the browser's binary.\cite{chromiumPins} Every time a user loads a pinned website, the certificate fingerprint is compared to the one provided in the binary. If the fingerprint matches, the client continues the SSL handshake. Otherwise, an error message is shown to the user explaining a secured connection couldn't be established.

Although this is a very efficient method to verify SSL certificates, it is difficult to deploy and maintain on a larger scale. Furthermore, the Chrome team needs to verify the "pin" definition manually before merging every pin request into the code branch. Therefore only larger websites have certificate pins.

\subsubsection{TACK}

TACK is a proposal by Moxie Marlinkspike and Trevor Perrin, providing a way to 'pin' TLS servers to the correct public key even when a Certificate Authority is delivering a different one. Although this is a promising proposition, it wouldn't protect against an attacker that might have a long-term MITM capability since pins are set on the first connection and only expire after some time.\cite{tackMITM}

\subsubsection{DANE}

The IETF proposal called \emph{DANE} is an attempt at large scale certificate pinning but by distributing the certificate fingerprint by DNS. This would enable website owners to specify their certificate fingerprint as a DNS entry. Visitors would then be able to verify the authenticity of the server.

Even if we consider that DNSSEC does provide good security, this system still relies on trusting the Domain Name System and its hierarchical structure.

\subsubsection{Tor Hidden Services}

Tor Hidden Services are reachable by hashes of public keys. This is of course the ideal case when it comes to security because the address itself contains information about the key. Unfortunately, humans are not good at remembering pseudorandom 16-character strings.

\subsection{Zooko's triangle}
In this section, we are covering Zooko's triangle conjecture.

\begin{figure}[h!]
\centering
\includegraphics[width=0.3\textwidth]{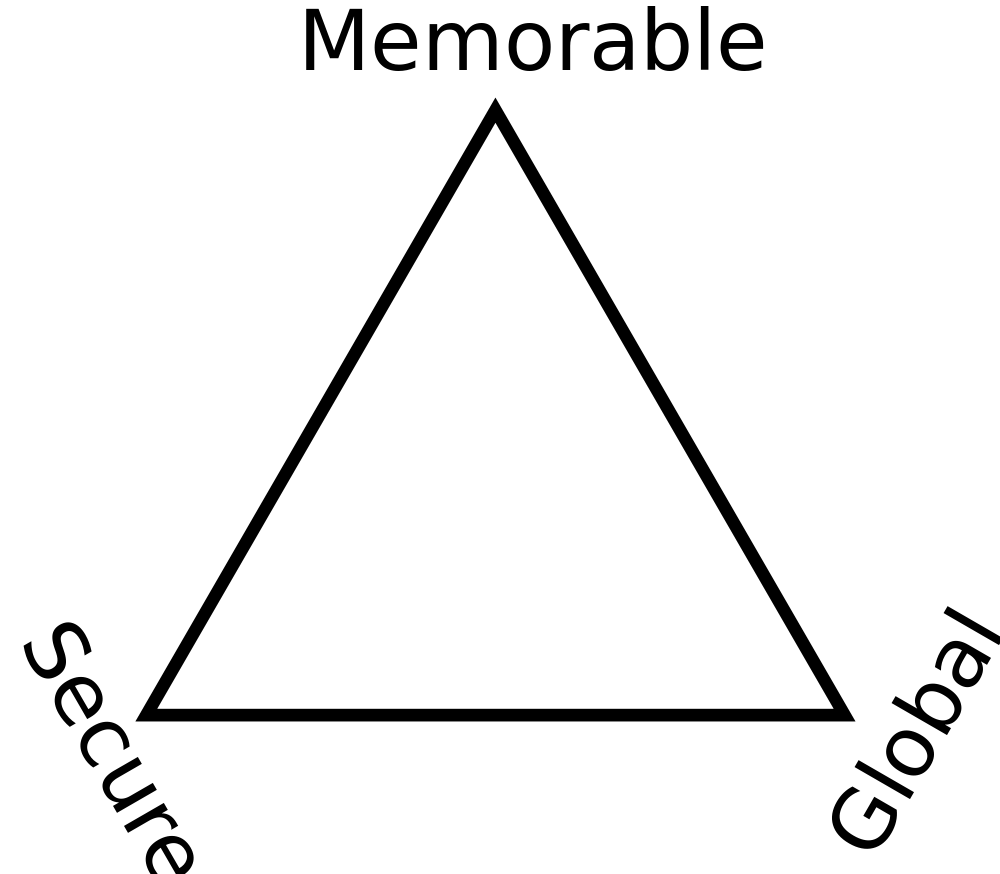}
\end{figure}

Zooko's triangle says that out of these three properties \cite{zookoTriangleWikipedia}, a naming system can only have two.
\begin{itemize}
\item \emph{Secure}\footnote{We cannot agree with the naming of this property given the threat model we described previously in this paper.}: The quality that there is one, unique and specific entity to which the name maps.
\item \emph{Global}: The lack of a centralized authority for determining the meaning of a name. Instead, measures such as a Web of trust are used.
\item \emph{Memorable}: The quality of meaningfulness and memorability to the users of the naming system.
\end{itemize}

We can thus see that the systems proposed so far only gather two out of three of those properties. If we take the DNSSEC system with DANE extensions, we can have a memorable address that is "secure". Unfortunately, this configuration does not have the global property because the ICANN is a centralized authority. Alternatively, Tor's Onion addresses do have the "secure" property and are global but a pseudorandom 16-character string is not memorable.

\subsection{Squaring Zooko's Triangle}

In the following section we will present a naming system that is an attempt at squaring Zooko's triangle.

Back in January 2011, Aaron Swartz described on his blog\cite{aaronBlog} how the Bitcoin blockchain could be of service in squaring Zooko's triangle. A few months later, a first implementation of that idea came into existence: Namecoin.

\subsubsection{The Bitcoin Blockchain}

The blockchain is Bitcoin's main innovation. Blockchains are mainly linear data-structures that were invented specifically for the Bitcoin project to store the history of all past transactions but they can be applied anywhere a distributed consensus needs to be established in the presence of malicious or untrustworthy actors.

To understand how they work we will cover the basics of how Bitcoin itself works. Let's take an example and see what happens when Alice tries to transfer money to Bob. Every user on the network has one address. Bitcoin addresses\cite{bitcoinSpec} are generated based on a public key. 
\begin{center}

Key-Hash = RIPEMD-160(SHA-256(public key))

$\text{BTC}_{\text{Address}}$ = Base58(Version +\footnote{The + sign is a string concatenation} Key-Hash + Checksum)
\end{center}

Private-public key pairs are generated\footnote{The curve used in Bitcoin is \emph{secp256k1} which is surprisingly a NIST recommended curve.} when creating a new Bitcoin addresses.

Alice must know Bob's address to send him money. Now that Alice has Bob's address, she creates a new message saying she sends a few Bitcoins to Bob and uses her private key to generate an ECDSA signature. Once she has generated that message and has signed it, she starts gossiping about her transaction on the network, her peers hear about the transaction, they verify if Alice has enough money to make the transaction and verify the signature. If the transaction looks legitimate, they start telling all of their peers. The verification can be done thanks to the blockchain data structure which is a decentralized and unique record of all the transactions. Peers that are miners, eventually hear about this transaction and add it to the transactions memory pool. This pool is a queue of transactions that are not yet merged in the blockchain. But now how can we merge transactions into the blockchain? 

\subsubsection{Proof of work}

The concept of \emph{proof of work} is used to merge the blockchain. It makes adding entries in the blockchain an expensive process computation-wise. Let's say Alice wants to send Bitcoins to Bob. Alice will start gossiping on the network, telling all her peers that she wants to send money to Bob. Every client, has a copy of the blockchain and can thus assess if Alice has the amount of money she wants to transfer to Bob. If she does, gossip will spread.

Once the miners, the workers of the blockchain, learn about a valid transaction (Alice has enough money to make the transaction and her signature is correct), they will add it to their memory pool. If the transaction is valid, the miners will add this transaction in the next block they will be mining. The benefit of making it costly to validate transactions is that validation can no longer be influenced by the number of network identities someone controls, but only by the total computational power they can bring to bear on validation.

So what is mining technically?

The hard challenge that is used in Bitcoin that needs to be solved is based on the strength of cryptographic hashes, also known as one-way functions. We consider that it is hard for someone to come up with the parameters of a hash functions for a given result. The function used in Bitcoin is \emph{SHA-256} but this hash function could be substituted by any other. Another cryptocurrency, Litecoin, chose to use the \emph{Scrypt} function.

If we want to add some blocks (list of transactions) to the blockchain, we will have to solve this problem
\begin{center}
SHA-256("TransactionsInfo" + challengeNumber) =< target
\end{center}
where \emph{transactionsInfo} is a parameter list of information about the transactions in the blocks (and some extra information like a return address for the reward). The blockchain is vulnerable to some malleability regarding certain informations in the transaction but all the important information (such as the amount of the transaction, the recipient and the sender) is part of this hash.
The challenge the miner has to solve --- the proof-of-work --- is to find the number such that when we append the transactions' infos to this number, and hash the combination, the output hash is smaller than a given number.
We notice that this certain challenge number is established by the network and determines how hard the problem is. In Bitcoin, this number is dynamically adjusted to keep an approximate block validation time of 10 minutes.\cite{hashCash}

When someone succeeds in solving this problem, they sends their solution to the network. Nodes verify if that answer is valid, and if it is, they broadcast it to their peers. It progressively spreads across all nodes and is added to the network's blockchain. 

\subsubsection{Dealing with collisions}

Now what happens if two nodes, from separate parts of the blockchain do succeed in solving the challenge at almost the same time. Both nodes and their peers will spread different versions of the blockchain. We say that the blockchain has \emph{forked}. How do we solve this?

In this case, miners will start mining the next block based on the version of the blockchain they have. If they hear that another blockchain is longer than the one they were working on before, they will switch to the longer one and put the transactions in the orphan blocks (blocks that were in the previous fork) back into the memory pool if they were not merged. 

We can now understand that because every node chooses to have the longest blockchain possible, it will be very hard for an attacker to spread a fake version of the blockchain because this would involve solving the challenge for every preceding block, since blocks are chained and must contain the block identifier of the previous one.

Why would one mine and spend so much computational power?

Miners are rewarded for their efforts. First, when making a transaction, we can speed up the money transfer by adding a transaction fee, that will go directly to the miners. Mining software is thus optimised to sort the transactions, in order to be merged in blocks by decreasing order of transaction fee. 
The other reward from mining comes from the coinbase transaction : mining does generate money.  At the creation of Bitcoin, this reward was set to be a 50 BTC. But for every 210,000 validated blocks (once every four years) the reward halves. This has happened just once, to date, and so the current reward for mining a block is 25 bitcoins. This halving in the rate will continue every four years until the year 2140 CE. At that point, the reward for mining will drop below $10^{-8}$ bitcoins per block which is a satoshi, the smallest unit of Bitcoin and the total amount of bitcoins will cease to increase.

\subsubsection{From Bitcoin to Namecoin}

Now that we understand how blockchains work and why they are safe data structures, let's now see how we can use them to square Zooko's triangle. 

Namecoin is a bitcoin fork that was designed as a decentralized key-value store in addition to a crypto-currency. Putting information in the blockchain does cost a certain price.

Namecoins can be spent in many ways, here are some other use cases of the Namecoin blockchain:
\begin{itemize}
\item Aliases: The blockchain can be used to store an easy to remember alias for a GPG/SSH key, a Bitcoin address or any other cryptographic identity.
\item Timestamping: The blockchain could store information about a specific file and from a hash of that we could find matching author name, owner, etc.
\item Messaging: The blockchain could be a decentralized store for  long-term messages vs BitMessage.
\end{itemize}

Writing data in the blockchain does have a certain price. Registering a domain does cost the registration fee (0.01NMC that goes to nobody) plus the transaction fee (that goes to the miner who succeeds in adding the block that contains this transaction).

The cost includes a network fee and a transaction fee. The fees are denominated in Namecoins (\emph{NMC}). Initially, the network fee was 50 NMC but it decreases twice every 2 months, which means that it is already less than 1 NMC after a year. This design was meant to make it expensive to register domains in the first few months to avoid the issue of domain name squatting.

Let's see what a domain name value message looks like to understand how it squares Zooko's triangle.

For a key \emph{d/\footnote{The d/ prefix is used to register a .bit domain} fredericjacobs}, we have 

\begin{lstlisting}[language=json,firstnumber=1]
{
    "ip"      : "209.236.123.133",
    "tor"     : "rqblqd3balaxcb57.onion",
    "email"   : "me@fredericjacobs.com",
    "info"    : "Frederic Jacobs",
    "tls": {
        "tcp": {
            443: [[1, "30F38EDAABC67F0344DBE27018552F7D575946EF", 1]]
        }
    },
    "map":
    {
        "www" : { "ip": "209.236.123.133" },
    }
}
\end{lstlisting}

This would make our website accessible at the \emph{fredericjacobs.bit}, a memorable, secure (a client can verify the fingerprint) and global address, the triangle has been squared!

A namecoin domain needs to be renewed every 36,000 blocks which at the current rate is around 200 days. Those updates are free. Hence, unlike ICANN domain names, you don't have to pay renewal fees.

\subsection{DNSNMC}

It might not be very convenient for users to require them to have a full copy of the blockchain on their client, especially if it's a mobile device. DNSNMC\cite{okTurtles} is a proposal for using the DNS client with Namecoin. Anyone can run a DNSNMC server\footnote{DNSNMC servers are not yet running but the software should be released in January 2014.} and configuring clients is easy since it's just a plug-and-play replacement for your DNS server. An DNSNMC client configuration consists of an IP address and a public key fingerprint to be able to verify that the DNS request was not modified. 

\subsection{Known Issues with this new model}

Blockchains are data-structures that are constantly growing. Given MinimaLT's keying requirements that we will describe later, the size of the blockchain grows substantially. Merkle trees require to know the hashes of the children nodes to verify integrity and our keying material obviously needs to be hashed to avoid malleability. Optimisations that could enable the miners to clean up the blockchain is still an open area of research.

Another issue that still needs to be addressed is domain squatting\cite{domainSquatting}. Because registering namecoin domains (.bit) became ridiculously cheap, a lot of domains are being squatted by people hoping to resell those domains at some point in the future. A better pricing system that prevents massive domain registration should be adopted because costs decreased too quickly to be an effective counter-measure.

\section{Transport security}

Now that we have a good long-term identity key distribution strategy, we are going to discuss a new transport-layer security protocol that provides safer and faster encrypted connections.
\subsection{MinimaLT}
MinimaLT\cite{MinimaLT} is a new\footnote{Paper published in May 2013} protocol that looks very promising.
It presents some of the interesting features we want. It was initially designed to secure network connections between computers running the secure operating system \emph{Ethos}. In the following sections we are going to compare it with TLS/TCP.\footnote{We will not however go over the whole specification of the protocol.}  Because MinimaLT is implemented on top of UDP, it is compatible with the current Internet infrastructure.

\subsection{Changes to make in Namecoin to make it play nicely with MinimaLT}

In the original MinimaLT paper, a directory service is introduced to deal with the exchange of hosts information. We will not cover that part but will rather explain how MinimaLT can be used along with the Namecoin blockchain.

The MinimaLT directories do store the following information about the hosts: IP Address, UDP port, long-term identity key\footnote{This key is not required in our use case where a client identifies to a server but can be useful for access control in other scenarios} and ephemeral key. We will see later in this paper why all these are needed, but let's first try to change the Namecoin domain JSON structure to add these new fields. 

Because we still want to be able to communicate over other protocols, we'll keep the base entry but add support for MinimaLT. 

\begin{lstlisting}[language=json,firstnumber=1]
{
    "ip"      : "209.236.123.133",
    "tor"     : "rqblqd3balaxcb57.onion",
    "email"   : "me@fredericjacobs.com",
    "info"    : "Frederic Jacobs",
    "tls": {
        "tcp": {
            443: [[1, "30F38EDAABC67F0344DBE27018552F7D575946EF", 1]]
        }
    },
    "minimaLT": {
        //"ip" : "209.236.123.133" // Only required if different than the default one
        "port": 80, // can be any UDP port.
        "id_key": long-term identity key,
        "eph_key": ephemeral key used for sending encrypted data on first RTT.
    }
    "map":
    {
        "www" : { "ip": "209.236.123.133" },
    }
}
\end{lstlisting}

For the perfect forward secrecy features of MinimaLT, we want new ephemeral keys to be announced regularly. Therefore, we also define a ephemeral key update message. 

\begin{lstlisting}[language=json,firstnumber=1]
{
    "minimaLT": {
        "eph_key": ephemeral key used for sending encrypted data on first RTT.
    }
}
\end{lstlisting}

Because the Namecoin blockchain is a key-value store, we do not need to specify what the address of our service is because it's the key of this entry. However, we do need to sign this ephemeral key update message with our Namecoin private key. 

\subsection{Why better than TLS/TCP?}

\subsubsection{Security is the standard}

Currently, on the Internet, most HTTP connections are unencrypted, and if they are, the Apache or Nginx configuration is often badly configured, in such a way that it could be attacked. MinimaLT was designed to prevent system administrators from making security mistakes. Never will you have to manually pick cipher suites or worry about OpenSSL compatibility, MinimaLT does it all for you. Not only does it encrypt all communications between two hosts but it defaults on strong ciphers\footnote{MinimaLT uses Curve25519 for the public key cryptography}.

\subsubsection{Encryption \& PFS}

A few months ago, a previously unknown email provider called \emph{Lavabit} was under pressure from the US government to hand over its private SSL key. The reason turned out to be that Edward Snowden was using the email service. If Lavabit had turned over the SSL keys, they would have compromised the privacy of their entire user base. This is due to poorly configured SSL where the ECDHE is not the default. ECDHE performs an Elliptic curve Diffie-Hellman key exchange. The last E from ECDHE stands for Ephemeral which means that a new key, that will never be stored, is generated at every handshake. Hence, even if the server gets compromised or the network operator is forced to hand over an encryption key, they won't be able to provide it. Unfortunately, most system administrators don't put the \emph{ECDHE} ciphers on top of their ciphers list resulting in the lack of Perfect Forward Secrecy.

\subsubsection{Tunnel oriented}

MinimaLT establishes tunnels. Tunnels provide a point-to-point encrypted channel to transmit information and have the interesting property of being more resistant to traffic analysis, just like IPSec. Tunnels mean that unlike TLS, all services from the transport layer, namely authentication, encryption, congestion control and reliability, are provided on a per-tunnel basis and are not repeated per connection. MinimaLT clients have one or more connections for each of these tunnels.

\subsubsection{IP Mobility}

Thanks to the structure of MinimaLT packets, the \emph{tunnel ID} is what identifies what packet belongs to what connection and therefore, MinimaLT has complete IP mobility. Unlike TCP, the source IP address and UDP port can change without affecting the connection. A specific RPC (\emph{$\text{nextTid}_{0}$}) exists to announce an IP address change. A change in IP address will cause a rekeying, a procedure that we describe further down. The fact that we have IP mobility is a big advantage over TCP that currently struggles in the mobile world when switching from a WiFi connection to a 3G/4G signal for instance. MinimaLT solves by design many problems that plague TCP today such as multi-path TCP!

\subsubsection{Handshakes \& Tunnel establishment}

When establishing a single TLS connection, two hosts must first go through a three-way TCP  handshake before they can start the TLS handshake which requires 4 more RTT. Thankfully MinimaLT attempts to fix that, and does perform, in most cases, a cryptographic handshake in \emph{less time than unencrypted TCP}! It's also important to note that handshakes with MinimaLT are way less frequent than in TLS because of the tunnel architecture.

Usually in TLS, the client needs to get the certificate first, before being able to send encrypted data, but with MinimaLT+Namecoin, we already have an ephemeral key that allows us to send encrypted data to the server. Let's see how this key exchange works.

Let's say a client \emph{C} wants to connect to a server \emph{S}. \emph{C} uses the Namecoin blockchain to get its identity, and an ephemeral key of \emph{S}. Now \emph{C} sends its first packet to \emph{S} containing its newly-generated ephemeral key, a new tunnel ID and the first bits of data. This segment's body can be encrypted using the ephemeral key of the server we retrieved previously ephemeral keys will no more be used past this point because both clients can now compute a shared secret by performing a Diffie-Hellman Key exchange with the public ephemeral keys they exchanged. In the future, this shared secret will be used to perform symmetric encryption between both hosts.

One important thing to notice is that this approach does provide perfect forward secrecy thanks to the Diffie-Hellman key exchange using ephemeral keys. Of course, perfect forward secrecy requires rekeying, ie the process of changing keys so that previously used keys can be cleared from memory. Let's go through this process. Rekeying can be requested by the client or the server depending on their rekey interval policies or change in IPs, but the client will always be the one initiating the rekey.

When a client initiates a rekey, it generates a new tunnel ID and sends it to the server as a call. Like the initial tunnel establishment procedure, the client generates a new key pair and sends its new public key along with the new tunnel ID to the server using the current encrypted tunnel. The corresponding private key of the pair that was generated will never be used (it actually doesn't even need to be known), its only function is to make the packet look like a regular tunnel initialisation packet. Now that the new tunnel is ready, the client can send packets to the new tunnel ID. But what encryption key should be used now? MinimaLT avoids doing a DH on rekeying for performance reasons, thus, client is hashing with a cryptographically secure hash function the old symmetric encryption key to generate the next one. The client will include the temporarily generated public key inside the encrypted packet sent to the server. Once the server gets a packet on the new tunnel ID, it will perform the same hashing as the client did to compute the decryption key. Now the server can decrypt the packet and it has to verify that the key is matching the previously sent one in the (\emph{$\text{nextTid}_{0}$}) call. This verification is required\cite{MinimaLT} because otherwise an active attacker could alter the key sent in the tunnel announce packet and then sent a second packet with the matching DH computer key and notice that it would fail. It could conclude that this would be a rekeying procedure and not the creation of a new tunnel.

\subsubsection{DDoS protection}

One might have legitimate concerns of how this would play out in a threat model were attackers would like to flood a system with the creation of tunnels because DH are expensive computationally-wise. One of the MinimaLT designers, Dan Bernstein, knows this problem really well because he invented the SYN cookies. In MinimaLT, puzzles are used to address the cases where servers are under load.

When a server is not experiencing any specific load, it does accept tunnel establishments requests without any questions asked but if the server is under load, it responds with a puzzle.

The puzzle requires a proof of work to be completed. The details of the proof of work that is used are not covered in this paper.

Benchmarking shows that even with a puzzle presented to the client, MinimaLT handshakes are faster than TLS/TCP and in most cases (without puzzles), they are faster than unencrypted TCP.\cite{MinimaLT}

Another way to exhaust the server's resources would be to flood him with data to decrypt, but given the fact that, past the key exchange, only symmetric key encryption is used, which on modern system is faster than the network links that would saturate faster.

\subsubsection{Congestion control}

MinimaLT's tunnel headers do contain fields for congestion control, such as sequence and acknowledgement numbers. MinimaLT doesn't provide anything additional to what exists in TCP for this task. MinimaLT replicates TCP's congestion control but does not provide anything beyond it.

\subsubsection{Flow control}

Every MinimaLT connection has its own receiving window size, similar to TCP. 

\section{MinimaLT and Anonymity}

MinimaLT has no anonymity network similar to the Tor Project yet. The Tor project has been investigating on using the SPEEDY protocol (basis of the HTTP2 specification) to speed up their connections between nodes. We can imagine that if MinimaLT gets mainstream adoption, the support of MinimaLT will be seriously considered.

\section{Conclusions}

We've seen how MinimaLT can help us speed up our encrypted connections and with the help of Namecoin, make them safer too. The model presented in this paper clearly has flaws, mainly with the scalability of the Namecoin blockchain and the issues it causes on mobile devices. Blockchains are only a few years old and many optimisations are still possible. They may not be perfect but they are the first data structure we know of that allows us to square Zooko's triangle! On the other hand, MinimaLT is a sneak peak into what tomorrow's transport-layer protocols will look like.

\balance

\section{Acknowledgments}
I would like to thank everyone who provided feedback and reviewed this paper. Particularly, 
\begin{itemize}
\item Greg Slepak who is currently working on the DNSNMC implementation and author of the OkTurtles\cite{okTurtles} paper for providing very valuable feedback to improve this paper.
\item All folks from Noisy Square at \#30c3 for the insightful discussions that influenced this paper.
\item Dylan Bourgeois, a friend of mine, for spotting so many typos.
\item Romain Ruetschi, another friend, for fixing a lot of my grammar mistakes.
\end{itemize}

\section{Ressources}

The source of this paper is available on GitHub along with some of the resources used to write this paper: \newline
\url{https://github.com/FredericJacobs/safeweb}

\bibliographystyle{abbrv}
\bibliography{bibliography}

\end{document}